\pdfoutput=1
\documentclass{bmcart}

\usepackage{amsthm,amsmath,amssymb}
\RequirePackage{hyperref}
\usepackage[utf8]{inputenc} 

\usepackage{graphicx} 

\startlocaldefs
\newcommand{\bra}[1]{\langle #1 \vert}
\newcommand{\ket}[1]{\vert#1 \rangle}
\newcommand{\Bra}[1]{\left\langle #1 \right\vert}
\newcommand{\Ket}[1]{\left\vert #1 \right\rangle}
\newcommand{\bBra}[1]{\Big\langle #1 \Big\vert}
\newcommand{\bKet}[1]{\Big\vert #1 \Big\rangle}
\newcommand{\KetBra}[2]{\Ket{#1}\kern-0.1em\Bra{#2}}
\newcommand{\Op}[1]{\ensuremath{\boldsymbol{\mathsf{\hat{#1}}}}}

\newcommand{\braket}[1]{\langle #1 \rangle}
\newcommand{\Avg}[1]{\left\langle #1 \right\rangle}
\newcommand{\Abs}[1]{\left|#1\right|}
\newcommand{\Norm}[1]{\lVert#1\rVert}
\newcommand{\diag}{\operatorname{diag}}
\newcommand{\tr}{\operatorname{tr}}
\newcommand{\hol}{\text{geo}}
\newcommand{\simplex}{\text{splx}}
\newcommand{\sm}{\text{sm}}
\newcommand{\Favg}{F_{\text{avg}}}
\newcommand{\EpsC}{\varepsilon_{C}}
\newcommand{\EpsPop}{\varepsilon_{\text{pop}}}
\newcommand{\EpsAvg}{\varepsilon_{\text{avg}}}
\renewcommand{\Im}{\operatorname{Im}}
\renewcommand{\Re}{\operatorname{Re}}
\endlocaldefs

\begin{document}

\begin{frontmatter}

\begin{fmbox}
\dochead{Research}


\title{Hybrid Optimization Schemes for Quantum Control}


\author[
   addressref={ks},                   
]{\inits{MG}\fnm{Michael H} \snm{Goerz}}
\author[
   addressref={bk},
]{\inits{KBW}\fnm{K. Birgitta} \snm{Whaley}}
\author[
   addressref={ks},
   email={christiane.koch@uni-kassel.de}
]{\inits{CK}\fnm{Christiane P} \snm{Koch}}


\address[id=ks]{
  \orgname{Theoretische Physik, Universität Kassel}, 
  \street{Heinrich-Plett-Str. 40},                     %
  \postcode{D-34132}                                
  \city{Kassel},                              
  \cny{Germany}                                    
}
\address[id=bk]{
  \orgname{Department of Chemistry}, 
  \street{University of California},                     %
  \city{Berkeley, CA},                              
  \postcode{94720}                                
  \cny{USA}                                    
}



\end{fmbox}


\begin{abstractbox}

\begin{abstract} 
Optimal control theory is a powerful tool for solving control problems in
quantum mechanics, ranging from the control of chemical reactions to
the implementation of gates in a quantum computer. Gradient-based optimization
methods are able to find high fidelity controls, but require considerable
numerical effort and often yield highly complex solutions. We propose here to
employ a two-stage optimization scheme to significantly speed up
convergence and achieve simpler controls. The control is initially
parametrized using only a few free parameters, such that optimization in this
pruned search space can be performed with a simplex method. The result,
considered now simply as an arbitrary function on a time grid, is the starting
point for further optimization with a gradient-based method that can quickly
converge to high fidelities. We illustrate the success of this hybrid technique
by optimizing a geometric phase gate for two superconducting transmon qubits
coupled with a shared transmission line resonator, showing that a combination of
Nelder-Mead simplex and Krotov's method yields considerably better results than
either one of the two methods alone.
\end{abstract}




\end{abstractbox}
%

\end{frontmatter}



\section{Introduction}

Coherent quantum control has long been a key component in the effort towards
future quantum technologies. It relies on the interference between multiple
pathways to steer the quantum system in some desired way~\cite{BrumerShapiro,
RiceZhao}.  Originally conceived for the control of chemical
reactions~\cite{TannorJCP1985, TannorJCP1986, LevinPRL15}, it has since been extended to
a wide variety of applications, see Ref.~\cite{BrifNJP2010} for a review.
In this context, numerical optimal control theory (OCT) is a particularly
powerful tool. OCT follows an iterative approach, improving the control field in
each iteration to steer the dynamics to the optimization target. Generally, the
fastest converging algorithms are those that take into account information about
the gradient of the optimization functional with respect to variations in the
control. Gradient-based methods assume an open-loop setup, where the
entire optimization procedure is performed based on the knowledge
of the system dynamics. Two widely used methods are Krotov's
method~\cite{KonnovARC99, SklarzPRA2002, ReichJCP12} and
gradient ascent pulse engineering (GRAPE)~\cite{KhanejaJMR05}.
Krotov's method guarantees monotonic convergence for
time-continuous control fields. The
Limited memory Broyden-Fletcher-Goldfarb-Shannon (LBFGS)
method~\cite{ByrdSJSC1995} can be used to extend both
GRAPE~\cite{FouquieresJMR2011} and Krotov's method~\cite{EitanPRA11}, considering
not only the gradient but also an estimate for the Hessian, i.e., the second
order derivative.
This has been demonstrated to improve the convergence in some instances, in
particular when close to the optimum~\cite{MachnesPRA2011}. In experimental
setups that only allow for limited control and knowledge of the dynamics,
closed-loop control schemes have often been preferred~\cite{JudsonPRL1992}.
There, the controls are updated based only on a measurement of the figure of
merit, e.g.\ using genetic algorithms~\cite{GoldbergGABook1989} or other
gradient-free optimization methods.

More recently, gradient-free optimization methods have also been employed in an
open-loop context, prompted by the observation that evaluation of the
gradient in many-body systems is often numerically infeasible. The
chopped random basis (CRAB) method~\cite{DoriaPRL11, CanevaPRA2011} has been
formulated for such applications. It expands the control in a relatively small
number of randomly chosen spectral components and then applies a Nelder-Mead simplex
optimization to the expansion coefficients. In principle, gradient-free methods
are applicable if the control can reasonably be described by only a few free
parameters and the optimization landscape has no local minima in the vicinity of
the initial ``guess''.

Optimal control theory is particularly relevant for quantum
information processing.  Both Krotov's method and GRAPE have been
extensively used
to obtain high-fidelity quantum gates~\cite{PalaoPRL02, TeschPRL2002,
TreutleinPRA2006, SchirmerJMO2009, MullerPRA11, EggerSST2014}. Short gate
durations are crucial, in order to minimize detrimental effects of decoherence.
With OCT, this is achieved by systematically decreasing the gate duration until
no solution can be found~\cite{CanevaPRL09, GoerzJPB11}, thus operating
at the quantum speed limit~\cite{BhattacharyyaJPA1983, MargolusPD1998}.
Moreover, OCT may be used to actively minimize the effects of
decoherence~\cite{KallushPRA06, Schulte-HerbruggenJPB2011}, and
to increase robustness with respect to classical noise~\cite{GoerzPRA2014}.
Robustness is a requirement that is generally difficult to fulfill with
analytical approaches.

Here, we explore the possibility of combining gradient-free and
gradient-based methods at different stages in the optimization, exploiting the
benefits of each method. The application of a simplex optimization to a guess
pulse, described by only a few free parameters, efficiently yields a comparably
simple first optimized pulse of moderate fidelity. This pre-optimized pulse
then  provides a good starting point for further optimization using
a gradient-based method. The second optimization stage relaxes the restrictions
on the search space implied by a simple parametrization and may then quickly
converge towards a high fidelity.  The simplex pre-optimization addresses the
observation that typically in direct gradient-based optimizations, due to the
large size of the search space, the majority of the numerical effort is spent in
``getting off the ground''.
Pre-optimization thus allows to locate a region of the search space in
which the gradient
is large enough to provide meaningful information.
Such good guess pulses may sometimes
be designed by hand, but this requires a very good intuition of the underlying
control mechanisms.
We propose here instead to simply prune the search space for the initial
phase of the optimization by reducing the complexity of the control.
The second optimization stage can then more easily identify high fidelity
solutions.

We illustrate the use of such hybrid optimization schemes by optimizing
a quantum gate on superconducting qubits, using an example inspired by the recently
proposed resonator-induced phase gate~\cite{CrossPRA2015}. Superconducting
circuits are a prime candidate for the implementation of quantum computing, due
to the flexibility in qubit parameters, their inherent
controllability, and the promise of scalability. Moreover, with recent
advances in the transmon architecture, decoherence times are approaching
0.1~ms~\cite{RigettiPRB2012}, allowing to reach fault-tolerance with
sufficiently fast gates.  However, the flexibility and large number of different
gate mechanisms~\cite{ChowNJP2013} also imply a challenge
from a control perspective, as it is not immediately obvious what are good qubit
parameters, or good guess pulses. This makes superconducting circuits especially
well-suited for combining a coarse search using simplex methods, which may then
be refined with a more powerful gradient-based method.

The paper is organized as follows: In section~\ref{sec:model}, we review the
realization of a geometric phase gate on a system of two transmon qubits coupled
with a shared transmission line resonator, using all microwave control.
In section~\ref{sec:direct}, we first introduce a functional targeting the
geometric phase gate, i.e., any diagonal perfect entangler, and then show the
results of a direct optimization using the gradient-based Krotov method. In
section~\ref{sec:hybrid}, we apply the gradient-free
Nelder-Mead simplex optimization to obtain pre-optimized guess pulses, which
then become the starting point for an optimization with Krotov's method. We
compare the control pulses obtained by the gradient-based, gradient-free, and
hybrid schemes, and the dynamics they induce, as well as the numerical effort
necessary to obtain converged results in each scheme.
Section~\ref{sec:conclusions} concludes.

\section{Model and Gate Mechanism}
\label{sec:model}


We consider a system of two transmon qubits, coupled via a shared transmission
line resonator (``cavity'')~\cite{JKochPRA07, MajerN2007}. The Hamiltonian reads
\begin{equation}
 \Op{H} =
       \sum_{q=1,2} \left[ \omega_{q} \Op{b}_{q}^{\dagger} \Op{b}_{q}
     + \frac{\alpha_q}{2} \Op{b}_{q}^{\dagger} \Op{b}_{q}^{\dagger} \Op{b}_{q}
     \Op{b}_{q}
     + g_q (\Op{b}_{q}^{\dagger} \Op{a} + \Op{b}_{q} \Op{a}^{\dagger})
     \right]
     + \omega_c \Op{a}^{\dagger}\Op{a}
     + \epsilon^*(t) \Op{a} + \epsilon(t) \Op{a}^{\dagger}\,,
     \label{eq:tm_fullham}
\end{equation}
where $\omega_c$, $\omega_1$, $\omega_2$ are the frequency of the cavity and the
first and second qubit, respectively; $\alpha_{1}$, $\alpha_{2}$ are the qubit
anharmonicities, and $g_{1}$, $g_{2}$ are the coupling between each qubit and
the cavity. The operators $\Op{a}$, $\Op{b}_1$, and $\Op{b}_2$ are the standard
annihilation operators for the cavity ($\Op{a}$) and the two qubits ($\Op{b}_1$,
and $\Op{b}_2$), respectively. For numerical purposes,
the Hilbert spaces for the qubit and cavity are truncated after 6, respectively
70, levels. It has been verified that the inclusion of additional levels
yields no significant change in the results of the subsequent sections.
The parameters take the values listed in Table~\ref{tab:parameters}. The system is driven by the microwave
field $\epsilon(t)$, with a pulse duration $T$. An off-resonant pulse results in
a state-dependent shift of the resonator frequency~\cite{CrossPRA2015}.
For a slowly-varying pulse shape with $\epsilon(0) = \epsilon(T) = 0$, such that
the level shifts occur adiabatically, the dynamics result in a geometric phase on
each of the qubit levels. That is, the resulting gate takes the diagonal
form
\begin{equation}
  \Op{U} = \diag\left[ e^{i \phi_{00}},  e^{i \phi_{01}},
                      e^{i \phi_{10}},  e^{i \phi_{11}} \right]\,.
  \label{eq:diag_gate}
\end{equation}

The maximal reachable concurrence of such a diagonal gate is
\begin{equation}
  C(\gamma) = \Abs{\sin\left(\frac{\gamma}{2}\right)}\,,
  \qquad
  \gamma \equiv \phi_{00} - \phi_{10} - \phi_{01} + \phi_{11}\,,
  \label{eq:concurrence}
\end{equation}
where $\gamma$ defines the non-local two-qubit phase~\cite{GoerzJPB11}.
The concurrence is obtained from the theory of local invariants for
two-qubit gates~\cite{MakhlinQIP2002,ZhangPRA03}. The local invariants for
a diagonal gate evaluate to $G_1 = \cos^2(\gamma/2)$ and
$G_2=1+2\cos^2(\gamma/2)$. From these, the Weyl chamber coordinates may be
calculated as $c_1 = \gamma/2$, $c_2 = c_3 = 0$~\cite{ZhangPRA03}.
Following Ref.~\cite{KrausPRA01}, the concurrence is evaluated as a function of
the Weyl chamber coordinates to yield Eq.~\eqref{eq:concurrence}. The gate is
a perfect entangler for $\gamma = \pi$.

We consider pulse shapes of the form
\begin{equation}
  \epsilon(t) = E_0 \sin^2 \left(\pi \frac{t}{T} \right) \cos(\omega_d t)
  \label{eq:pulse_sin_sq}
\end{equation}
with a fixed driving frequency $\omega_d$ given in
Table~\ref{tab:parameters}.
For simplicity, we neglect the dephasing induced by high cavity populations
which allows solution of wave-packet dynamics with the time-dependent
Schrödinger equation including microwave control fields.
For an arbitrarily chosen
guess pulse of duration $T = 200$~ns and peak amplitude $E_0 = 300$~MHz,
the population dynamics resulting from the
initial condition $\Ket{\Psi(t=0)} = \Ket{00}$ are shown in panels
(a)--(d) of Fig.~\ref{fig:octJhol_popdyn}.

If the pulse induces adiabatic dynamics, it shifts the
qubit and cavity levels proportionally to $\epsilon(t)^2 / \Delta$ where
$\Delta$ is the detuning from the respective level~\cite{CrossPRA2015}.
In the original field-free frame, this is equivalent to shifting the initial
wave packet proportionally to the square of the pulse. Thus, the excitation and
population dynamics should smoothly follow the pulse shape. Specifically,
the condition for adiabaticity is that if the pulse shape were to be stretched
in time, the population dynamics would simply stretch correspondingly.
For a given peak amplitude, the larger the detuning of the drive
$\omega_d$ from the frequencies $\omega_c$, $\omega_2$, $\omega_1$, the
smaller the excitation in the respective Hilbert space. Since the drive is
detuned by only 40~MHz from the cavity, the cavity excitation, panel (a),
reaches a large value $\Avg{n} \approx 30$. The far-detuned right qubit, panel (b),  and
even farther detuned left qubit, panel (c), only show a small excitation. As
indicated by the standard deviations shown as shaded areas in panels (a)--(c),
the excitation remains relatively localized
in energy. It is noteworthy, however, that the
excitation curves (specifically of the cavity) show some imperfections. These
``wobbles'' can be interpreted as deviations from the expected adiabatic
dynamics, e.g.\ jumping over an avoided crossing between highly excited
cavity states. This ultimately results in a small loss of population from
the logical subspace, as the system does not perfectly return to its original
state. The loss of population for the given example is $\EpsPop = 5.9 \times
10^{-3}$. While such a small deviation is not discernible in the plot of the
population in panel (d), it is nonetheless a significant error in the objective
of obtaining high fidelity gates below the quantum error correction limit,
typically resulting in gate errors below $10^{-3}$.
In principle, cavity population can be suppressed by tuning the qubit
parameters and using non-trivial pulse shapes~\cite{CrossPRA2015}.

The dynamics for the remaining two-qubit basis states are similar to those of
the $\Ket{00}$ state in Fig~\ref{fig:octJhol_popdyn}. The concurrence of the gate
implemented by the propagated guess pulse, according to
Eq.~\eqref{eq:concurrence},
yields a value of $C \approx 0.8$. This implements the closest diagonal perfect
entangler with an average gate error of $8.3 \times 10^{-2}$.
 Both the loss of population from the
logical subspace and the small value of the generated entanglement imply that
the chosen pulse parameters are sub-optimal with respect to the desired
geometric phase gate. We therefore turn to numerical optimal control
to obtain a high fidelity gate.

\section{Direct Optimization with Krotov's Method}
\label{sec:direct}

\subsection{Optimization Functionals and Method}
\label{subsec:functionals}

The standard approach for implementing a specific quantum gate $\Op{O}$ using
optimal control theory is to maximize the overlap between the time evolution
$\Op{U}(T,0;\epsilon(t))$ under the control $\epsilon(t)$  and the target
gate~\cite{PalaoPRL02}.
For a two-qubit gate, this is commonly expressed in the final time
functional~\cite{PalaoPRA03}
\begin{equation}
  J_{T}^{\sm}
  = 1 - \frac{1}{16} \Abs{\sum_{k=1}^{4} \bBra{k} \Op{O}^{\dagger}
                                          \Op{U}(T,0;\epsilon(t)) \bKet{k}}^{2}\,,
  \quad
  \Ket{k} \in \left\{ \Ket{00}, \Ket{01}, \Ket{10}, \Ket{11} \right\}\,,
  \label{eq:J_T_sm}
\end{equation}
which goes to zero as the target gate $\Op{O}$ is implemented, up to a global
phase.

Krotov's method allows to iteratively improve the control field, changing the
control $\epsilon^{(i)}(t)$ to the updated control $\epsilon^{(i+1)}(t)$ in the
$i$'th iteration. The final time functional $J_T$, given e.g.\ by
Eq.~\eqref{eq:J_T_sm}, is augmented with a running cost to result in the total
functional
\begin{equation}
  J[\epsilon^{(i)}(t)]
  = J_T(\{\phi_k^{(i)}(T)\})
    + \int_0^{T} g_a[\epsilon^{(i)}(t)] \, dt\,,\quad
\ket{\phi_k^{(i)}(t)} = \Op{U}(t,0;\epsilon^{(i)}(t)) \Ket{k}\,.
\end{equation}
Monotonic convergence is ensured by the choice~\cite{PalaoPRA03,ReichJCP12}
\begin{equation}
  g_a[\epsilon^{(i)}(t)]
  = \frac{\lambda_a}{S(t)} (\Delta \epsilon(t))^2\,,
  \quad
  \Delta \epsilon(t) \equiv \epsilon^{(i+1)}(t) - \epsilon^{(i)}(t)\,,
\end{equation}
where $\lambda_a$ is an arbitrary scaling parameter and $S(t) \in [0,1]$ is
a shape function that ensures smooth switch-on and switch-off.

The iterative update scheme is then given in terms of three coupled
equations~\cite{ReichJCP12},
\begin{subequations}
\label{eq:krotov_coupled}
\begin{eqnarray}
  \Delta\epsilon(t)
  &=& \frac{S(t)}{\lambda_a} \Im \Bigg[
      \sum_{k=1}^{4} \Bigg(
      \Bigg\langle \chi_k^{(i)}(t) \Bigg\vert
        \bigg( \frac{\partial \Op{H}}{\partial \epsilon}
              \bigg\vert_{%
              \substack{\phi^{(i+1)}(t) \\ \epsilon^{(i+1)}(t)}}
        \bigg)
      \Bigg\vert \phi_k^{(i+1)}(t) \Bigg\rangle
      \nonumber \\ && \quad
      + \frac{1}{2} \sigma(t)
      \Bigg\langle \Delta\phi_k(t) \Bigg\vert
        \bigg( \frac{\partial \Op{H}}{\partial \epsilon}
              \bigg\vert_{%
              \substack{\phi^{(i+1)}(t) \\ \epsilon^{(i+1)}(t)}}
        \bigg)
      \Bigg\vert \phi_k^{(i+1)(t)} \Bigg\rangle
    \Bigg)\Bigg]\,, \label{eq:krotov_update}
  \\
  \frac{\partial}{\partial t} \Ket{\phi_k^{(i+1)}(t)}
  &=& - \frac{i}{\hbar} \Op{H}^{(i+1)} \Ket{\phi_k^{(i+1)}(t)},
    \quad
    \Ket{\phi_k^{(i+1)}(0)} = \Ket{k}\,, \label{eq:krotov_fw}
  \\
  \frac{\partial}{\partial t} \Ket{\chi_k^{(i)}(t)}
  &=& - \frac{i}{\hbar} \Op{H}^{\dagger (i)} \Ket{\chi_k^{(i)}(t)},
    \quad
    \Ket{\chi_k^{(i)}(T)}
     = - \frac{\partial J_T}{\partial \bra{\phi_k}} \bigg\vert_{\phi_k^{(i)}(T)}\,,
     \label{eq:krotov_bw}
\end{eqnarray}
\end{subequations}
with $\ket{\Delta\phi_k(t)} \equiv \ket{\phi_k^{(i+1)}(t)} - \ket{\phi_k^{(i)}(t)}$.
The second order contribution to the update, with the prefactor $\sigma(t)$, is
required for certain types of functionals~\cite{ReichJCP12}, as we
will see below. For the choice
of $J_{T}^{\sm}$, we may set $\sigma(t) = 0$~\cite{ReichJCP12}.

For the gate mechanism outlined in Section~\ref{sec:model}, we obtain
a phase on each of the four logical basis states. These phases should
combine to produce a perfect entangler, with $\gamma = \pi$ according to
Eq.~\eqref{eq:concurrence}. The individual phases $\phi_{00}$, $\phi_{01}$,
$\phi_{10}$ and $\phi_{11}$ depend delicately on the shape, amplitude, and
duration of the pulse. It is therefore not known \textit{a priori} for a given guess pulse
which exact geometric phase gate will or can be reached, and thus what should be
the target gate $\Op{O}$ of the optimization. For a gate $\Op{U}_0$ induced by
a guess pulse, we may construct the \emph{closest} diagonal perfect entangler
by numerically
evaluating
\begin{equation}
  \Op{O}
  = \underset{\phi_{00}, \phi_{01}, \phi_{10}}{\arg \min}
    \Norm{\Op{O}_{\diag}(\phi_{00}, \phi_{01}, \phi_{10}) - \Op{U}_0}\,,
\label{eq:closest_holonomic}
\end{equation}
with
\begin{equation}
  \Op{O}_{\diag}(\phi_{00}, \phi_{01}, \phi_{10})
  = \diag\left[ e^{i \phi_{00}},  e^{i \phi_{01}}, e^{i \phi_{10}},
            e^{i (\pi + \phi_{01} + \phi_{10} - \phi_{00})} \right]\,,
\end{equation}
which includes the condition $\gamma = \pi$ to make the gate a perfect
entangler. Using this gate as a target fully determines the optimization
problem, with two caveats.
First, the target gate depends on $\Op{U}_0$ induced by the
(arbitrary) guess pulse, and second, the construction of the closest diagonal
perfect entangler does not take into account the topology of the
optimization landscape; the ``closest'' gate is by no means guaranteed to be the
one that is easiest to reach.

An approach that addresses these issues is to go beyond the standard functional
of Eq.~\eqref{eq:J_T_sm} and formulate a functional that targets the
properties of the geometric phase gate specifically, without stipulating the
phases on all of the logical states. We split the functional into two terms,
\begin{equation}
  J_{T}^{\hol} = \frac{1}{8} ( J_{\diag} + J_{\gamma} )\,,
  \label{eq:J_T_hol}
\end{equation}
where $J_{\diag}$ goes to zero if and only if the gate is
diagonal (with arbitrary phases) and $J_{\gamma}$ goes to zero if and only if
the gate is also a perfect entangler, $\gamma = \pi$. Thus, the functional is
conceptually similar to a recently proposed functional targeting an arbitrary
perfect entangler~\cite{PE1}. However, the additional restriction to enforce
a diagonal gate is important, as the Hamiltonian also allows for non-diagonal
gates, but only through undesired non-adiabatic effects.

The two terms take the form
\begin{subequations}
\begin{eqnarray}
  J_{\diag} &=& 4 - \tau_{00}\tau_{00}^* - \tau_{01}\tau_{01}^*
          - \tau_{10}\tau_{10}^* - \tau_{11}\tau_{11}^*\,,\\
  J_{\gamma} &=& 2 + \tau_{00} \tau_{01}^* \tau_{10}^* \tau_{11}
                 + \tau_{00}^* \tau_{01} \tau_{10} \tau_{11}^*\,,
  \label{eq:J_gamma}
\end{eqnarray}
with
\begin{equation}
 \tau_{00} \equiv \braket{00|\Op{U}|00}\,, \quad
 \tau_{01} \equiv \braket{01|\Op{U}|01}\,, \quad
 \tau_{10} \equiv \braket{10|\Op{U}|10}\,, \quad
 \tau_{11} \equiv \braket{11|\Op{U}|11}\,.
\end{equation}
\end{subequations}
The construction of $J_{\gamma}$ is based on the observation that
\begin{equation*}
  \gamma = \pi
  \quad \Longleftrightarrow \quad
  2 + e^{i \gamma} + e^{-i \gamma} = 0\,,
\end{equation*}
which becomes Eq.~\eqref{eq:J_gamma} by associating $\tau_k$ with $e^{i
\phi_{k}}$ and using the definition of $\gamma$ in Eq.~\eqref{eq:concurrence}.
Both $J_{\diag}$ and $J_{\gamma}$ take
values $\in [0, 4]$, hence the normalization factor $\frac{1}{8}$ in
Eq.~\eqref{eq:J_T_hol} to bring the value of the functional closer to that of
$J_{T}^{\sm}$.

In contrast to $J_T^{\sm}$, the functional $J_T^{\hol}$ is \emph{not}
convex, since the states enter in higher than quadratic order. The Krotov update
equation~\eqref{eq:krotov_update} must then include the second order
contribution, where $\sigma(t)$ can be determined numerically in each iteration
as~\cite{ReichJCP12}
\begin{equation}
  \sigma(t) =  - \max(\epsilon_A, 2A + \epsilon_A)\,,
  \quad
  A  =
  \frac{2 \sum_{k=1}^{4} \Re\left[\langle \chi_k(T) \vert \Delta\phi_k(T) \rangle \right]
        + \Delta J_T}
       {\sum_{k=1}^{4} \Abs{\Delta\phi_k(T)}^2}\,,
\end{equation}
with a small non-negative number $\epsilon_A$,
and $\Delta J_T \equiv J_T(\{\phi_k^{(i+1)}(T)\}) -J_T(\{\phi_k^{(i)}(T)\})$.
The boundary condition for the backward propagated states in
Eq.~\eqref{eq:krotov_bw} yields
\begin{subequations}
\begin{eqnarray}
  \Ket{\chi_{00}(T)}
  &=& \left( \tau_{00} - \tau_{01}\tau_{10}\tau_{11}^* \right)
     \Ket{00}, \\
  \Ket{\chi_{01}(T)}
  &=& \left( \tau_{01} - \tau_{00}\tau_{10}^*\tau_{11} \right)
     \Ket{01}, \\
  \Ket{\chi_{10}(T)}
  &=& \left( \tau_{10} - \tau_{00}\tau_{01}^*\tau_{11} \right)
     \Ket{10}, \\
  \Ket{\chi_{11}(T)}
  &=& \left( \tau_{11} - \tau_{00}^*\tau_{01}\tau_{10} \right)
     \Ket{11}.
\end{eqnarray}
\end{subequations}

Both $J_{T}^{\sm}$ and $J_{T}^{\hol}$ are only
loosely connected to the \emph{average gate fidelity} that is accessible to
experimental measurement. In the case of a two-qubit gate and non-dissipative
dynamics this can be evaluated as~\cite{PedersenPLA2007}
\begin{equation}
  \Favg = \int \Abs{\bBra{\Psi} \Op{O}^\dagger \Op{U} \bKet{\Psi}}^2 d\Psi
        = \frac{1}{20} \left(
            \Abs{\tr\left[\Op{O}^\dagger \Op{U}\right]}^2
            + \tr\left[ \Op{O}^\dagger \Op{U} \Op{U}^\dagger \Op{O}\right]
          \right)\,.
\end{equation}
Thus, $\Favg$, respectively the gate error $\EpsAvg \equiv 1-\Favg$, provides a well-defined
measure of the optimization success independent of the choice of
optimization functional.
For an optimization with
$J_{T}^{\hol}$, we may evaluate $\EpsAvg$ with respect to the closest geometric
phase gate resulting from propagation with the optimized pulse, according
to Eq.~\eqref{eq:closest_holonomic}.

\subsection{Optimization Results}
\label{subsec:oct_hol}


The optimization starts from the  guess pulse described by
Eq.~\eqref{eq:pulse_sin_sq}, with $T = 200$~ns and $E_0 = 300$~MHz, as discussed
in Section~\ref{sec:model}, with the dynamics shown in panels (a)--(d) of
Fig.~\ref{fig:octJhol_popdyn}. The gate error with respect to the
closest geometric phase gate for this guess is $\EpsAvg = 8.3 \times 10^{-2}$, with
a loss of population from the logical subspace of $\EpsPop = 5.9 \times
10^{-3}$.  The concurrence error, defined as $\EpsC \equiv 1 - C$, takes the
value $1.9 \times 10^{-1}$.


Optimization using Krotov's method and $J_{T}^{\hol}$ as the optimization
functional converges within 5516 iterations of the algorithm. Convergence is
assumed when the relative change of the functional $\Delta J_{T} / J_{T}$ falls
below $10^{-4}$, such that no significant further improvement is to be expected.
The gate error is reduced to $\EpsAvg = 1.4 \times 10^{-4}$. It is dominated by
the remaining loss of population from the logical subspace,
$\EpsPop \approx \EpsAvg$, as the concurrence
error is only $\EpsC = 1.8 \times 10^{-6}$, see Table~\ref{tab:comparison}.

The resulting optimized pulse is shown in Fig.~\ref{fig:octJhol_pulse}, with the
guess pulse indicated by the dashed line. For the center 100$\,$ns, there are only
small deviations from the guess pulse (both in shape and phase).
Significant deviations occur only at the very beginning and end, most notably the
variation in the complex phase (center panel) between 20 and 40$\,$ns. The spectral
width of the pulse (bottom panel) remains well within a bandwidth of $\pm
50\,$MHz. The dynamics of the $\Ket{00}$ state under the optimized
pulse are shown in panels (e)--(h) of Fig.~\ref{fig:octJhol_popdyn}. The
difference to the dynamics under the guess pulse, panels (a)--(d), is striking;
the excitations no longer smoothly follow the pulse shape, but show strong
oscillations on top of the expected behavior. The features at the beginning of
the optimized pulse provide a kick to the system,
inducing oscillations in the populations,
with a counter-kick near the end of the pulse. These kicks are very
visible in the population of the $\ket{00}$ state in panel (h), compared to the
smooth dynamics for guess pulse in panel (d).

It is worth noting that the gate obtained with the optimized pulse is
\emph{not} the closest geometric phase gate to the gate implemented by the guess
pulse, as in Eq.~\eqref{eq:closest_holonomic}. This illustrates the
benefit of using $J_T^{\hol}$ over $J_T^{\sm}$. The latter optimizes towards
a specific, pre-determined gate, according to Eq.~\eqref{eq:closest_holonomic},
while $J_T^{\hol}$ can dynamically adjust which specific geometric phase gate
is easiest to reach, allowing it to fulfill the objective much more easily.
Table~\ref{tab:comparison} shows that optimization with $J_T^{\hol}$ requires
significantly less propagations (which are directly proportional to CPU
time) than optimization with $J_T^{\sm}$, for both direct and pre-optimized
strategies.

While the optimization yields a gate error well below the quantum error correction
threshold, it deviates significantly from the simple geometric phase gate scheme,
resulting in complex dynamics. The numerical effort required to obtain a high
fidelity solution is considerable, with several thousand iterations (each
iteration requiring two full propagations of four logical basis states).
We have discussed here only the optimization for a fixed gate duration of
$T = 200\,$ns.
Generally, significantly faster quantum gates could
potentially be implemented using other mechanisms, e.g.~\cite{EggerSST2014}.
The geometric phase gate, however, relies on adiabatic shifts of the
energy levels such that
loss of population from the logical subspace inhibits realization of
high fidelities when pushing the gate duration significantly below 200$\,$ns.

The complexity of the optimized pulse is typical for Krotov's method
or other gradient-based optimization methods. For the present example, this
clashes with the gate mechanism of the geometric phase gate that intends to use
simple and smooth pulse shapes.

\section{Hybrid Optimization Scheme}
\label{sec:hybrid}

\subsection{Simplex Optimization}
\label{subsec:simplex}

The gate error of the guess pulse is dominated by the insufficient amount of
entanglement that is generated. A natural approach is to maintain the analytical
pulse shape of Eq.~\eqref{eq:pulse_sin_sq} for the time being, and to vary the
free parameters $E_0$ and $T$ in order to maximize the figure of merit. Such an
optimization of a pulse determined by only a handful of parameters (in this case
two) is easily performed using a gradient-free method such as Nelder-Mead
simplex. This has the additional benefit that there are no restrictions on the
choice of optimization functional. Specifically, there is no need to formulate
it in such a way that derivatives can be calculated analytically. Thus, we can
include the objective of minimizing the gate duration in the functional and
modify Eq.~\eqref{eq:J_T_hol} to read
\begin{equation}
  J_{T}^{\simplex} = J_{\diag} + J_{\gamma} + \frac{T}{T_0}\,,
  \quad
  T_0 = 200\,\text{ns.}
  \label{eq:J_T_simplex}
\end{equation}
Note that with the addition of penalizing the gate duration, the functional
is no longer a distance measure that approaches zero as the target is reached;
the simplex optimization will find a local minimum of Eq.~\eqref{eq:J_T_simplex}
at a value of $J_{T}^{\simplex}\lesssim 1$.


Using only 116 propagations, the algorithm converges to a solution that
reduces the gate duration from 200 to 185$\,$ns, while bringing the entanglement
error down to $\EpsC = 2.0 \times 10^{-5}$, see table~\ref{tab:comparison}.
The resulting dynamics are shown in panels (a)--(d) of
Fig.~\ref{fig:simplex_popdyn}. They are similar to those of the original guess
pulse, cf.\ panel (a)--(d) of Fig.~\ref{fig:octJhol_popdyn}. The shorter pulse
duration and larger pulse intensity (from 300$\,$MHz in the original guess to
$\approx 400\,$MHz) results in a significantly larger cavity excitation. It also
leads to more non-adiabatic defects (wobbles in the cavity excitation).
Consequently, the loss of population from the logical subspace is increased by
about a factor of two to $\EpsPop = 1.4 \times 10^{-2}$, and limits the total
gate error to $\EpsAvg = 1.4 \times 10^{-2}$. Thus, while the simplex search
yields a dramatic improvement over the original guess pulse, it does not reach
a sufficiently high fidelity to approach the quantum error correction limit. To
remedy this, we turn to a hybrid approach, combining simplex and gradient-based
optimization.

\subsection{Continued Optimization with Krotov's Method}
\label{subsec:continued}

For the present example, we use the final pulse
of the previous section as the starting point of an optimization with Krotov's
method. Since now the guess pulse already has a relatively high fidelity,
both the $J_T^{\sm}$ and $J_T^{\hol}$ functionals may be used interchangeably,
as we are only searching in a very small vicinity of the starting point.
Both methods converge rapidly to $\Delta J_T/J_T < 10^{-4}$ in under 200
iterations. The dynamics resulting from the propagation of the $\ket{00}$ state
under the pulse obtained from optimization with the $J_T^{\sm}$ functional is
shown in panels (e)--(h) of Fig.~\ref{fig:simplex_popdyn}. The comparison to the
dynamics of the pre-optimized guess pulse in panels (a)--(d) is striking: the
excitations now follow the pulse shape smoothly. The non-adiabatic defects,
i.e., the wobbles especially in the cavity excitation in panel (a), have
been corrected. Consequently, the loss of population from the logical subspace
is now reduced to a value of $\EpsPop = 1.1 \times 10^{-5}$.
Together with only a slight increase in the concurrence error to
$\EpsC = 5.1 \times 10^{-5}$, the overall gate error of the optimized pulse is
$\EpsAvg = 3.4\times 10^{-5}$. This is an improvement of half an order of
magnitude compared to the direct optimization in section~\ref{subsec:oct_hol}.
Moreover, the result has been obtained at a small fraction of the numerical cost.
The pre-optimized guess and post-optimized pulse shape, indicated as the blue
shaded area in panels (d) and (h), appear visually indistinguishable.


The correction to the pre-optimized guess pulse is shown in
Fig~\ref{fig:postJsm_diff}. Indeed, the corrections are on the order of 1$\,$MHz,
much smaller than the peak amplitude of $E_0 \approx 400\,$MHz. The corrections
follow a regular pattern, both in the shape (top panel) and the complex phase
(center panel), as indicated by the existence of sharp peaks in the spectrum
(bottom panel). It appears that the non-adiabatic defects in the pre-optimized
guess pulse can be corrected by a series of small kicks in amplitude and phase
at regular intervals. This results in an optimized pulse that is conceptually
simpler and yields a higher fidelity than the direct application of Krotov's
method. The comparison illustrates the power of a hybrid approach to steer the
\emph{physical} characteristics of the optimized control. In our example, the
parametrization used in the first optimization stage enforces the desired
pulse shape. Without this restriction, pulses of undesirable
complexity are obtained. On the other hand, the restriction must ultimately be
relaxed to allow for the representation of the higher-frequency features that
correct non-adiabatic defects. Thus, the role of the two-stage optimization is
also to influence the physical features of the control, in addition to its
numerical benefits.

The striking numerical efficiency of the hybrid optimization scheme can be seen
by comparing its convergence to that of the direct optimization. This is shown
in Fig.~\ref{fig:convergence}. The direct optimization shows an extended plateau
for at least the first 100 iterations, before slowly converging. This behavior is
typical for ill-chosen guess pulses~\cite{GoerzNJP2014}. For a direct
optimization with the $J_T^{\sm}$ functional, the plateau extends for several
thousand iterations, and does not yield convergence within 10000 iterations.
In contrast, the optimization starting from a pre-optimized pulse has no plateau
(note the log-scale of the x-axis); both $J_T^{\sm}$ and $J_T^{\hol}$ converge
at roughly the same rate. The improvement by the simplex (pre-)optimization
compared to the original guess pulse can be seen from the difference in the
y-intercept between the ``direct'' and ``pre-optimized'' curves
in Fig.~\ref{fig:convergence}; in addition,
there is also a reduction in the gate duration from 200 to 185$\,$ns
at no additional numerical cost.

The gate duration for the geometric phase gate mechanism is limited by the
requirement of adiabaticity. We have also performed the hybrid optimization
scheme for a gate duration of $\approx100$~ns. In this case, the simplex search
yields significant non-adiabatic defects, with a loss of population of $1.2
\times 10^{-1}$. The concurrence error is $2.3\times 10^{-3}$ and the total gate
error, $\EpsAvg = 1.2 \times 10^{-1}$, is dominated by the population loss.
Post-optimization using Krotov's method significantly reduces the total gate
error to $\EpsAvg = 1.4 \times 10^{-2}$. The post-optimization result is
obtained at a numerical cost very close to that for the $T=200$~ns gate, and
yields pulse corrections very similar to those shown in
Fig.~\ref{fig:postJsm_diff}. This results in a correction of the non-adiabatic
defects and thereby lowers the population loss to only $6.5\times 10^{-3}$. The
total gate error is now dominated instead by the increased concurrence error of
$1.7\times 10^{-2}$, which is insufficient for a high quality phase gate.  The
observation that an overall improvement in gate performance at $T = 100$~ns from
hybrid optimization is only possible by increasing adiabaticity at the cost of
reduced entanglement, indicates that a quantum speed limit has been reached for
the specific gate mechanism.  These results show that a hybrid optimization
scheme may be used successfully even when operating close to the quantum speed
limit.

While it is not the aim of this work to characterize the quantum speed limit for
the coupled transmon system, we note that this could be quantified by
systematically scanning over the gate duration~\cite{GoerzJPB11}. Under the
constraints of the specific gate mechanism employed here, a numerical approach
appears necessary to extract the speed limit. However, for the Strauch gate
of Ref.~\cite{StrauchPRL2003}, an analytic estimate may be made from summing
the minimal precession periods required for each component of the gate, i.e.,
two iSWAP and one controlled-Z gates. With the qubit-cavity interaction strength
of $g = 70$~MHz employed here, this would suggest that gate durations as short
as a few tens of nanoseconds might be attainable for this system.


We may also compare the total number of propagations necessary to obtain the
optimized pulse, included in Table~\ref{tab:comparison} together with measures
of success: concurrence error, loss of population from the logical subspace, and gate
error. Each iteration using Krotov's method requires two propagations.
For the hybrid optimization schemes (``pre-opt s.m.'', ``pre-opt hol.''), the
number of propagations includes simplex as well as the additional propagations
due to Krotov's method. The gate error is measured with respect to the exact
target gate when using the square-modulus functional, and with respect to the
closest geometric phase gate to the optimized dynamics in all other cases. In terms of
numerical efficiency, the hybrid
schemes outperform the direct optimization by nearly two orders of magnitude
while resulting in a significantly better gate fidelity.

\section{Conclusions}
\label{sec:conclusions}

For the example of a geometric phase gate on a system of two
transmon qubits with a shared transmission line resonator, we have considered
the application of gradient-based optimization methods, specifically Krotov's
method, and a gradient-free optimization method, Nelder-Mead simplex. The
objectives of the geometric phase gate can be formulated in a specialized
optimization functional $J_T^{\hol}$ that reaches its optimal value for any
diagonal perfect entangler. We have shown that for a direct
optimization, this functional vastly outperforms the ``standard'' square-modulus
functional for gate optimization.  Convergence is aided by the
use of a functional that formulates the objective as general as possible. This
is in agreement with recent results for optimization using a functional
targeting arbitrary perfect entanglers~\cite{PE1, PE2}.

The direct optimization using Krotov's method can in principle find controls
that implement the desired gate with high fidelity. However, the resulting
dynamics are complex and the numerical effort is dominated by an extended
plateau in the initial phase of the optimization; for short gate durations,
optimization becomes increasingly harder and the fidelity is limited by loss
from the logical subspace. The numerical effort to reach high fidelities quickly
becomes unfeasible in this case.
In contrast, parametrizing the pulse by its duration and peak amplitude only,
and applying a simplex optimization on those parameters, we are able to find
a simple analytic pulse that implements the desired gate with moderate
fidelity, still one order of magnitude above the quantum error
correction limit. Thus, for the example presented here, neither the application
of a gradient-based algorithm nor the simplex optimization alone yield
satisfactory results; only the combination of both methods into a hybrid
optimization scheme is able to obtain controls with a clear mechanism that
implement a geometric phase gate to high fidelity, with a minimum of numerical effort.

These results prompt the recommendation to generally adapt hybrid optimization
schemes, i.e., obtain guess pulses for gradient-based optimization from
a gradient-free pre-optimization, when there is insufficient knowledge to design
good guess pulses by hand. There is great flexibility in the choice of
parametrization. Here, we have taken the two free parameters, peak
amplitude and gate duration, in a simple fixed analytical formula for the
pulse shape. Generally, one might use slightly more sophisticated
parametrizations, following e.g.\ the CRAB approach~\cite{CanevaPRA2011}.
The small number of free parameters and relatively high quality of the original
guess pulse with a fidelity of already $>90\%$ results in particularly fast
convergence of the simplex method. For a more sophisticated example, at least
several hundred propagations would probably be required in the simplex stage.
However, even in that case, this numerical effort is by far outweighed
by the large number of iterations
required to leave the initial plateau in the optimization landscape for a bad
guess pulse in a direct gradient-based optimization.
Moreover, any figure of merit is suitable for optimization with the simplex
method, as there is no need to derive the gradient for the
optimization functional. For example, the gate duration can be included in the
figure of merit, something that generally is not straightforward in gradient-based
methods~\cite{NdongJMO14}. The hybrid scheme is aimed at providing optimal solutions in
an open-loop context, and still leaves open the possibility of further combining
open-loop and closed-loop optimization methods when targeting a specific
experimental setup~\cite{EggerPRL2014}.

In principle, the approach could thus be extended to multiple stages, where in
each stage, a different parameterization and a suitable optimization algorithm
is used. There is no requirement for a specific method such as Nelder-Mead
simplex or Krotov's method to be employed.  For example, a two-stage
optimization using a genetic algorithm in the first stage, and a gradient
algorithm in the second stage has been used for the optimization of quantum
gates in strongly coupled two-level systems~\cite{GraceJPB2007, GraceJMO2007}.
There, starting with a simple parametrization, and then moving to a less
restricted search space in the second optimization stage was also found to
benefit the overall optimization performance, in agreement with the results
shown here.

Adding an additional stage might also be beneficial when the optimization
landscape is non-trivial and contains traps or saddle points. This may happen
e.g.\ when the optimization is performed with limited
resources~\cite{MooreJCP2012}. Repeating the simplex search from a different
starting point---either by systematic variation or by random search---may then
find solutions that do not get stuck in traps of the optimization landscape. Of
course, the truncation of the search space due to the low-dimensional
parametrization may itself introduce additional traps in the landscape. However,
these traps would disappear when returning to the full search space in the final
optimization stage.

For the specific example of the geometric phase gate considered here, we have found
the post-optimization to introduce small corrections to non-adiabatic defects.
Due to the small relative strength of the correction, the experimental
realization of the geometric phase gate would
require extraordinary precision. Moreover, the large excitation of the cavity
would limit the fidelity when dissipative effects, specifically spontaneous
decay of the cavity are taken into account. However, we stress that the method
of employing hybrid optimization schemes presented here is entirely general and
aimed at reducing the numerical effort in obtaining high fidelity solutions to
arbitrary quantum control problems, not limited to quantum information
processing. Moreover, robustness with respect to both fluctuations in the
control parameters and dissipation can be achieved using complimentary advanced
control techniques, such as a description in Liouville space and ensemble
optimization~\cite{GoerzNJP2014, GoerzPRA2014}. These approaches are
numerically even more demanding, such that the reduction of
the optimization cost achieved by the hybrid scheme discussed here may
become imperative.


\begin{backmatter}

\section*{Competing interests}
The authors declare that they have no competing interests.

\section*{Author's contributions}
MHG has constructed the model and optimization framework under the supervision
of KBW and CPK; he carried out all numerical calculations. All authors
discussed the results. MHG wrote the manuscript with contributions from KBW
and CPK.

\section*{Acknowledgements}
We would like to thank F.~Motzoi for fruitful discussions. Financial support
from DAAD under grant PPP USA 54367416 is gratefully acknowledged.


\bibliographystyle{bmc-mathphys} 
\bibliography{preopt}





\section*{Figures}
\clearpage

  \begin{figure}[h!]
  \includegraphics{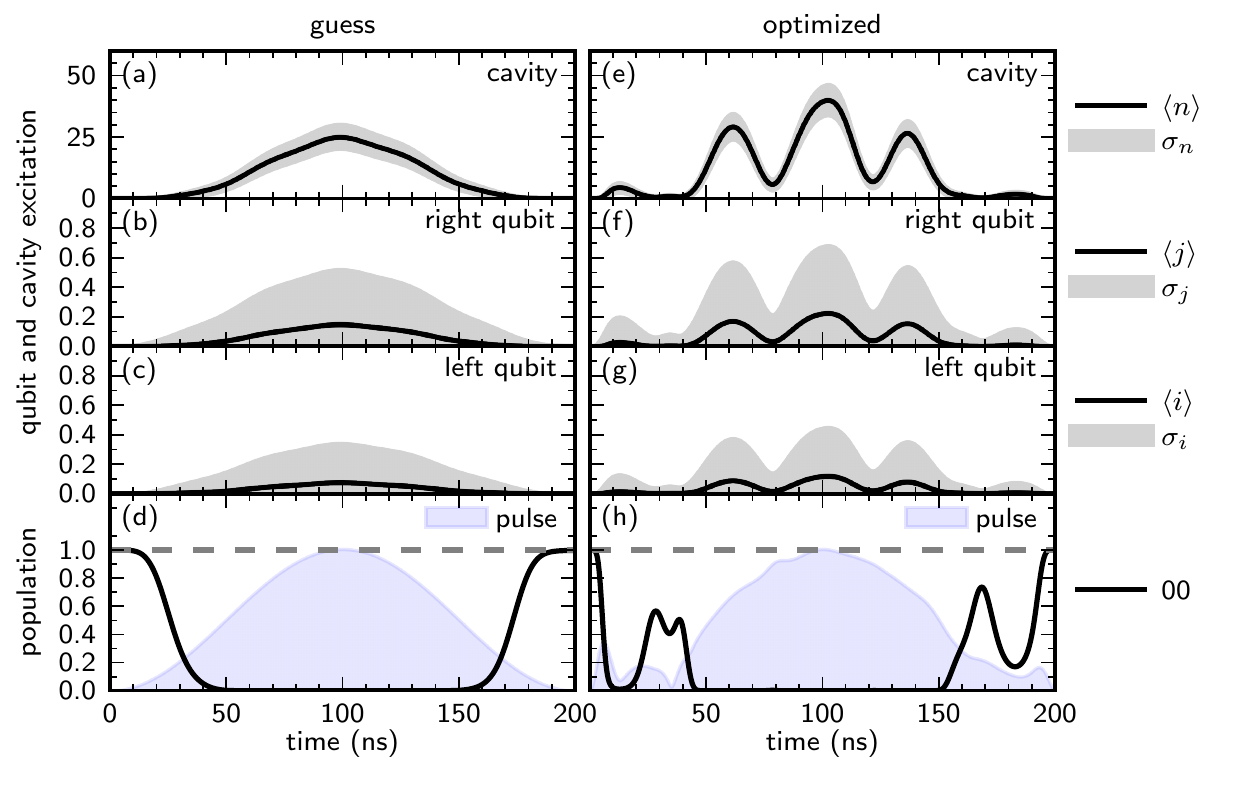}
  \caption{\csentence{Population dynamics under guess and optimized pulse.}
  The figure shows the population dynamics of the initial state
  $\Ket{\Psi}(t=0) = \Ket{00}$ for the guess pulse, panels (a)--(d), and the
  pulse obtained from direct optimization of the $J_{T}^{\hol}$ functional,
  panels (e)--(h). In panel (a), (e), expectation value $\Avg{n}$ of the cavity
  excitation, plus-minus the standard deviation $\sigma_n$. In panels (b), (f),
  and (c),(g), expectation values and standard deviations for the excitation of
  the right and left qubit, respectively.  In panel (d), (h), population in the
  state $\Ket{00}$. The pulse shape (normalized by the peak amplitude
  $E_0 =$~300~MHz, cf.~Fig.~\ref{fig:octJhol_pulse}) is shown in
  the background of panels (d), (h).
  The guess pulse implements a geometric phase gate with a gate error of
  $\EpsAvg = \text{8.3} \times \text{10}^{\text{-2}}$,
  with concurrence error
  $\EpsC = \text{1.9} \times \text{10}^{\text{-1}}$ and
  population loss from the logical subspace
  $\EpsPop = \text{5.9} \times \text{10}^{\text{-3}}$.
  The optimized pulse decreases the gate error to
  $\EpsAvg = \text{1.4} \times \text{10}^{\text{-4}}$, with
  $\EpsC = \text{1.8} \times \text{10}^{\text{-6}}$ and
  $\EpsPop = \text{1.4} \times \text{10}^{\text{-4}}$, see
  Table~\ref{tab:comparison}.
  }
  \label{fig:octJhol_popdyn}
  \end{figure}

  \begin{figure}[h!]
  \includegraphics{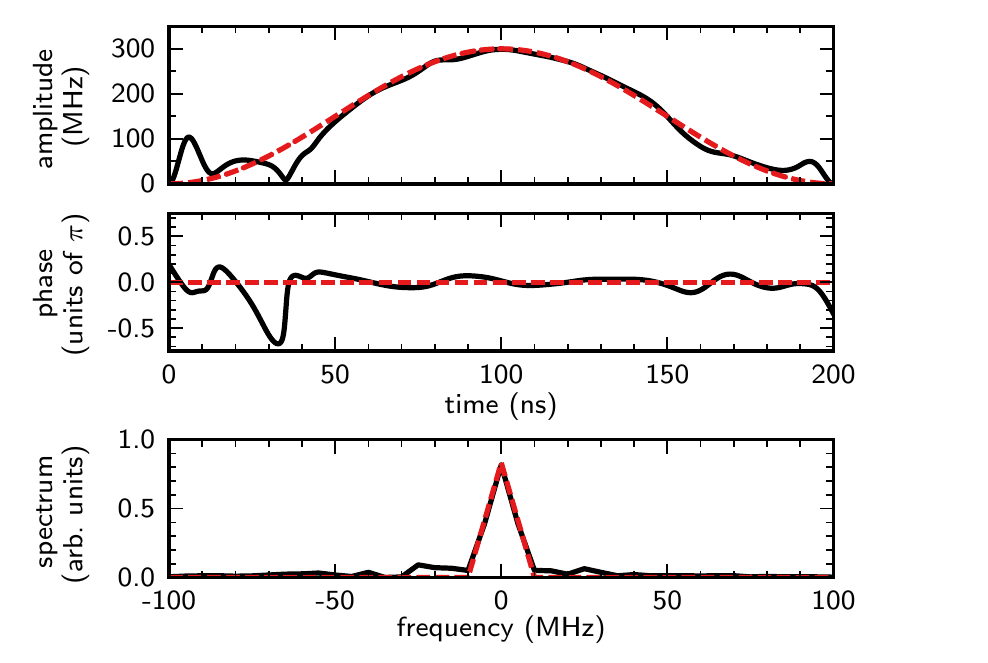}
  \caption{\csentence{Optimized pulse resulting from direct
      optimization with Krotov's method using the
  $J_{T}^{\hol}$ functional.} In the panels from top to bottom, absolute value
  of the pulse shape, complex phase, and spectrum of the optimized pulse (solid
  black lines) and of the guess pulse (dashed red/gray lines).}
  \label{fig:octJhol_pulse}
  \end{figure}

  \begin{figure}[h!]
  \includegraphics{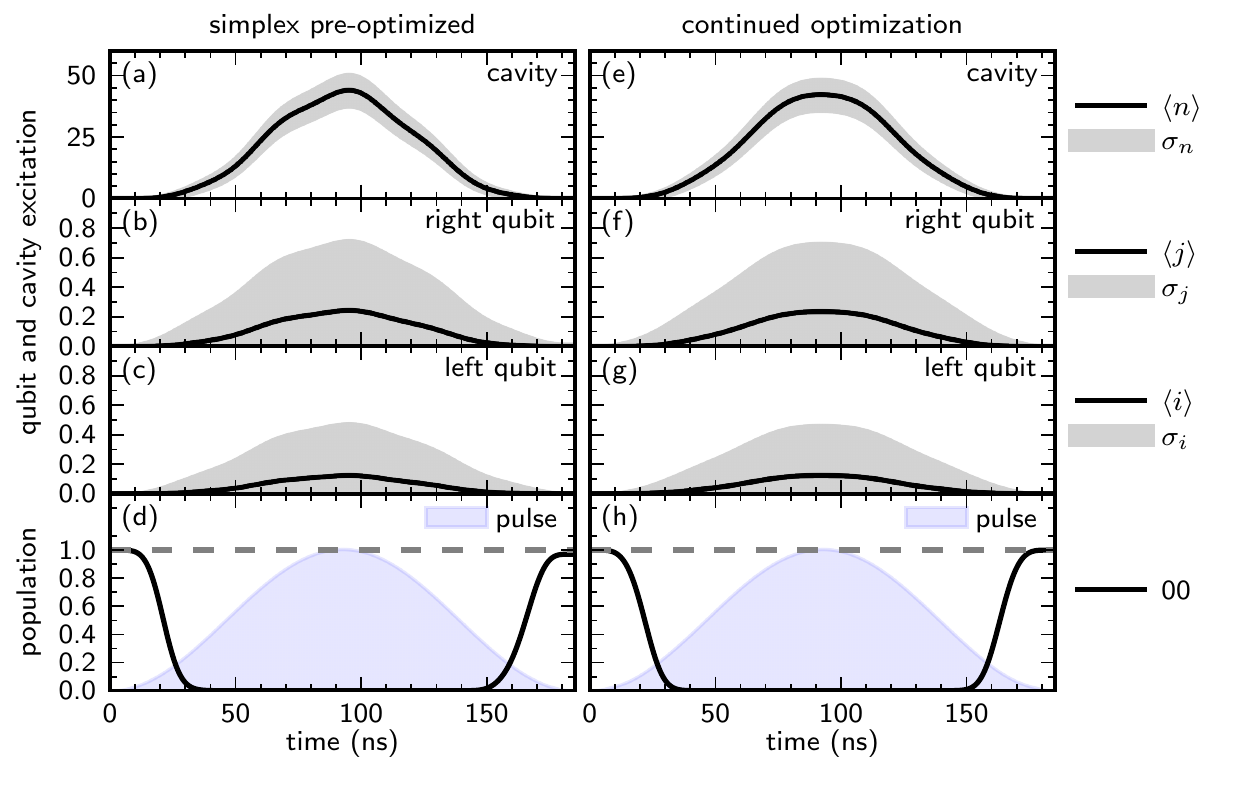}
  \caption{\csentence{Population Dynamics under (pre-optimized) guess and
  optimized pulse.}
  The figure follows the conventions of Fig.~\ref{fig:octJhol_popdyn}.
  Panels (a)--(d) show the dynamics resulting from a simplex optimization, see
  text for details. The resulting pulse is the
  starting point for a continued optimization using Krotov's method with the
  $J_{T}^{\sm}$ functional. The optimized dynamics are shown in panels (e)--(h).
  The pulse amplitude indicated in the background of panels (d), (h) is
  normalized to the peak amplitude of $E_0 \approx$~400~MHz.
  The simplex-optimized pulse implements a geometric phase gate with a gate error of
  $\EpsAvg = \text{1.4} \times \text{10}^{\text{-2}}$, with
  $\EpsC = \text{2.0} \times \text{10}^{\text{-2}}$ and
  $\EpsPop = \text{1.4} \times \text{10}^{\text{-2}}$.
  The continued optimization decreases the gate error to
  $\EpsAvg = \text{3.4} \times \text{10}^{\text{-5}}$, with
  $\EpsC = \text{5.1} \times \text{10}^{\text{-5}}$ and
  $\EpsPop = \text{1.1} \times \text{10}^{\text{-5}}$, see
  Table~\ref{tab:comparison}.
  }
  \label{fig:simplex_popdyn}
  \end{figure}

  \begin{figure}[h!]
  \includegraphics{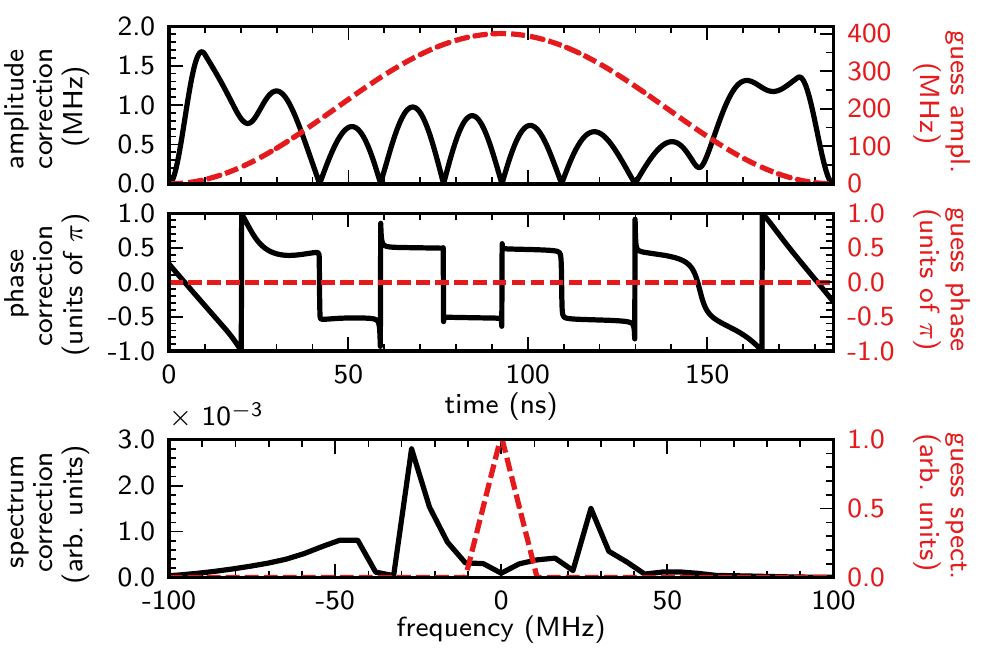}
  \caption{\csentence{Pulse corrections obtained with Krotov's method.}
  The figure summarizes the differences between the optimized pulse, cf.~panel~(h) in
  Fig.~\ref{fig:simplex_popdyn}, and the (pre-optimized) guess pulse,
  cf.~panel~(d) in Fig.~\ref{fig:simplex_popdyn}, also indicated by the dashed
  red/gray line in each panel. The panels from top to bottom
  show the corrections to absolute value, complex phase, and spectrum of the
  pulse, respectively (solid black lines). The amplitude, phase, and
  spectrum of the guess pulse (dashed red/gray lines) are shown using the
  alternative axis scaling in red/gray.
  }
  \label{fig:postJsm_diff}
  \end{figure}

  \begin{figure}[h!]
  \includegraphics{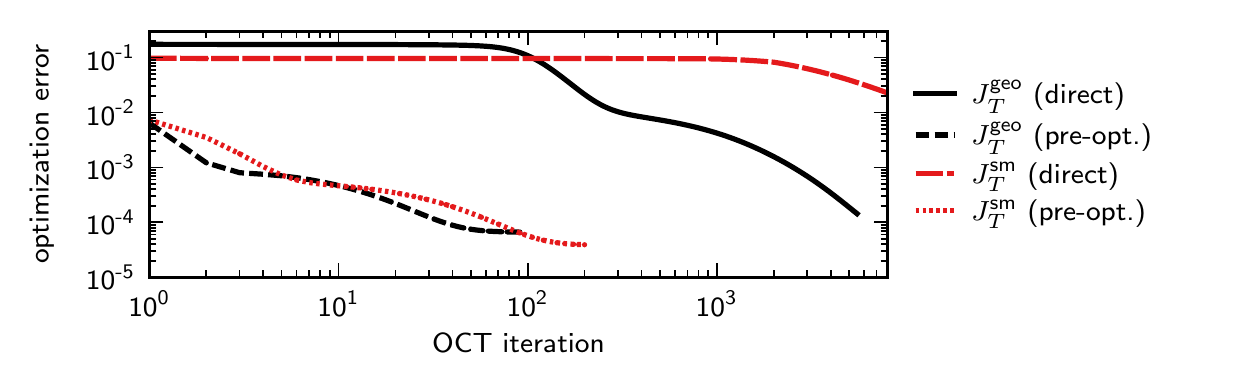}
  \caption{\csentence{Convergence of optimization towards a geometric phase gate.}
  Value of the final-time optimization functional $J_T^{\hol}$, respectively
  $J_T^{\sm}$, over the number of iterations using Krotov's method. Each
  iteration requires two full propagations; the number of propagations are
  proportional to the required CPU time. The direct
  optimization starts from an arbitrary guess pulse, see
  section~\ref{subsec:oct_hol} for details. In the pre-optimized case, the guess
  pulse was the result of a simplex optimization, see
  section~\ref{subsec:continued} for details.
  }
  \label{fig:convergence}
  \end{figure}



\begin{table}[h!]
\caption{Parameters for two transmon qubits coupled via a shared transmission
line resonator}
  \centering
  \begin{tabular}{llrl}
  \hline
  cavity frequency      &  $\omega_c$            & 8.10       &GHz \\
  left qubit frequency  &  $\omega_1$            & 6.85       &GHz \\
  right qubit frequency &  $\omega_2$            & 7.25       &GHz \\
  drive frequency       &  $\omega_d$            & 8.14       &GHz \\
  qubit anharmonicity   &  $\alpha_1, \alpha_2$  & -300       &MHz \\
  qubit-cavity coupling &  $g_1, g_2$            &   70       &MHz \\
  \hline
  \end{tabular}
  \label{tab:parameters}
\end{table}

\begin{table}[h!]
\caption{Optimization success for different optimization schemes. For each
scheme, we give the gate duration $T$, the total number of propagations, the
concurrence error $\EpsC \equiv 1 - C$ by which the gate
differs from a perfect entangler, the loss of population $\EpsPop$ from the
logical subspace, and the gate error $\EpsAvg \equiv 1 - \Favg$ with respect to
a geometric phase gate. The number of propagations for ``pre-opt. s.m.'' and
``pre-opt. geo.'' include both the 116 propagations of the first stage
simplex optimization and the propagations from the second-stage optimization
using Krotov's method, with 201 respectively 92 iterations, and two propagations
per iteration. The reported number of propagations is thus proportional to the
total CPU and wall-clock time required to obtain the result starting from the
original guess pulse.
}
  \begin{tabular}{lrrrrr}
  \hline
  {} &  $T$ [ns] &  prop. &                $\EpsC$     &            $\EpsPop$  &             $\EpsAvg$ \\
  \hline
  guess         &       200 &      0 & $\text{1.92} \times \text{10}^{\text{-1}}$ & $\text{5.94} \times \text{10}^{\text{-3}}$ & $\text{8.25} \times \text{10}^{\text{-2}}$ \\
  direct s.m.   &       200 &  20000 & $\text{2.23} \times \text{10}^{\text{-2}}$ & $\text{4.13} \times \text{10}^{\text{-3}}$ & $\text{1.45} \times \text{10}^{\text{-2}}$ \\
  direct geo.   &       200 &  11032 & $\text{1.83} \times \text{10}^{\text{-6}}$ & $\text{1.42} \times \text{10}^{\text{-4}}$ & $\text{1.43} \times \text{10}^{\text{-4}}$ \\
  simplex       &       185 &    116 & $\text{1.95} \times \text{10}^{\text{-5}}$ & $\text{1.40} \times \text{10}^{\text{-2}}$ & $\text{1.40} \times \text{10}^{\text{-2}}$ \\
  pre-opt. s.m. &       185 &    518 & $\text{5.07} \times \text{10}^{\text{-5}}$ & $\text{1.11} \times \text{10}^{\text{-5}}$ & $\text{3.36} \times \text{10}^{\text{-5}}$ \\
  pre-opt. geo. &       185 &    300 & $\text{5.24} \times \text{10}^{\text{-5}}$ & $\text{1.40} \times \text{10}^{\text{-5}}$ & $\text{3.50} \times \text{10}^{\text{-5}}$ \\
  \hline
  \end{tabular}
  \label{tab:comparison}
\end{table}


%

\end{backmatter}
\end{document}